\documentclass[manuscript,screen]{acmart}
\renewcommand\footnotetextcopyrightpermission[1]{} 
\AtBeginDocument{%
  \providecommand\BibTeX{{%
    \normalfont B\kern-0.5em{\scshape i\kern-0.25em b}\kern-0.8em\TeX}}}
\usepackage{graphicx}
\usepackage{amsmath}
\usepackage{tcolorbox}
\usepackage{listings}
\usepackage{multirow}
\usepackage{pdflscape}
\usepackage{hyperref}
\usepackage{lmodern}
\usepackage{MnSymbol}
\usepackage{stmaryrd}
\usepackage{shuffle}

\usepackage{amsthm}
\usepackage{booktabs}

\usepackage{tikz}
\usepackage{lineno}
\usepackage{pdfpages}
\usepackage{multicol}
\usepackage{listings}
\usepackage{enumitem}

\newlist{steps}{enumerate}{1}
\setlist[steps, 1]{label = Step \arabic*:}
\usetikzlibrary{arrows.meta,
                chains,
                positioning,
                shapes.geometric
                }

\setcopyright{none}
\renewcommand\footnotetextcopyrightpermission[1]{}
\begin{document}
\nolinenumbers
\title{\textit{e}GEN: An Energy-saving Modeling Language and Code Generator for Location-sensing of Mobile Apps}

\author{Kowndinya Boyalakuntla}
\affiliation{%
  \institution{Indian Institute of Technology Tirupati}
  \streetaddress{Settipalli Post}
  \city{Tirupati – 517 506}
  \country{India}}
\email{cs17b032@iittp.ac.in}
\orcid{0000-0002-3112-9718}

\author{Marimuthu C}
\affiliation{%
  \institution{National Institute of Technology Karnataka}
  \streetaddress{Surathkal}
  \city{Mangalore}
  \country{India}}
\email{cs15fv08.muthu@nitk.edu.in}
\orcid{0000-0002-4905-0530}

\author{Sridhar Chimalakonda}
\affiliation{%
  \institution{Indian Institute of Technology Tirupati}
  \streetaddress{Settipalli Post}
  \city{Tirupati – 517 506}
  \country{India}}
\email{ch@iittp.ac.in}
\orcid{0000-0003-0818-8178}

\author{K. Chandrasekaran}
\affiliation{%
  \institution{National Institute of Technology Karnataka}
  \streetaddress{Surathkal}
  \city{Mangalore - 575 025}
  \country{India}}
\email{kchnitk@ieee.org}

\renewcommand{\shortauthors}{Kowndinya and Marimuthu, et al.}

\begin{abstract}
The demand for reducing the energy consumption of location-based applications has increased in recent years.
The abnormal battery-draining behavior of GPS makes it difficult for the developers to decide on battery optimization during the development phase directly. 
It will reduce the burden on developers if battery-saving strategies are considered early, and relevant battery-aware code is generated from the design phase artifacts.
Therefore, we aim to develop tool support, \textit{e}GEN, to specify and create native location-based mobile apps.
\textit{e}GEN consists of Domain-specific Modeling Language (DSML) and a code generator for location-sensing. 
It is developed using Xtext and Xtend as an Eclipse plug-in, and currently, it supports native Android apps. 
\textit{e}GEN is evaluated through controlled experiments by instrumenting the generated code in five location-based open-source Android applications. 
The experimental results show 4.35 minutes of average GPS reduction per hour and 188 mA of average reduction in battery consumption while showing only 97 meters degrade in location accuracy over 3 kilometers of a cycling path.
Hence, we believe that code generated by \textit{e}GEN would help developers to balance between energy and accuracy requirements of location-based applications. The source code, documentation, tool demo video\footnote{\url{https://youtu.be/J-ZmzEBpC8Y}}, and tool installation video\footnote{Part1 of installation at \url{https://youtu.be/wyDfAoAlP-c} and part2 of installation at \url{https://youtu.be/o65Fu-xlByI}}are available at \url{https://github.com/Kowndinya2000/egen}. 
\end{abstract}



\keywords{domain-specific language, code generator, energy-saving location-sensing, modeling adaptive strategies, mobile apps}

\maketitle
\section{Introduction}\label{S:introduction}
In recent years, software energy consumption is becoming a critical non-functional requirement for software developers \cite{pang2016programmers,linares2018multi,ciancarini2020analysis}. The investigations by \citet{10.1145/3154384},  and \citet{manotas2016empirical} highlight the significance of reducing software energy consumption.
Especially in the mobile application domain, reducing the abnormal energy consumption is a concern for both users and developers. \citet{cruz2017performance} have performed an empirical evaluation on six popular android applications namely \textit{Loop - Habit Tracker}, \textit{GnuCash} and so on to find performance-related code smells that cause heavy battery depletion. They reported that smartphone's battery life can be extended upto an hour, if apps are developed by implementing energy-aware practices. 
This motivates the need for further research towards making energy efficient mobile applications.
Research efforts were initially focused on 
energy measurement \cite{hao2013estimating}, energy profiling \cite{hoque2015modeling}, and energy bugs identification \cite{Pathak:2012:KMP:2307636.2307661, liu2014greendroid}. Of late, the focus has been shifted to energy-saving solutions \cite{mcintosh2019can, schuler2020characterizing}, and automated repairing of energy bugs \cite{banerjee2018energypatch, morales2018earmo, cruz2018using}. Recently, the research community has provided several frameworks \cite{linares2018multi, CANETE2020106220, pereira2021ranking} to reduce energy consumption of Android apps. Notably, the research work by \citet{10.1145/3337773} highlights the list of techniques and tools to adopt in the software development life cycle to improve energy efficiency. 

The energy bugs are induced in the mobile domain by the usage of energy-hungry components such as GPS, GPU, camera, gyroscope, and other smartphone sensors \cite{pathak2012keeping}. Notably, the energy consumed by location-based apps may drain the mobile battery quickly if GPS is not appropriately handled by the deveopers\cite{capurso2017android}. 
\citet{huang2018location} discussed about the transformative evolution of location based services in mobile applications ranging from navigation, health care and social networking to assistive technologies and disaster management. 

The ubiquitous presence of location-based apps makes the creation of GPS efficient patterns even more challenging\cite{dutta2018energy, chen2019modeling}. To ease out developers' challenges in reducing GPS consumption, a potential strategy is to consider energy-saving decisions during the design phase of software development. In addition, it is important to enable provision for developers to consider energy-saving decisions independent of the location sensing libraries and code patterns.

\color{black}
The existing approaches primarily consider energy-saving solutions at source code level~\cite{pinto2017energy, ribeiro2021ecoandroid,cruz2017performance,cruz2017leafactor}. However, they do not consider energy-saving strategies during early stages of software development such as design, which might help developers handle energy-related problems. Hence, an appropriate tool support to consider energy-saving solutions at design time could help developers handle energy-related problems more clearly. In addition, automatically generating suitable battery-aware code from design phase artifacts might increase the developer's productivity. To the best of our knowledge, tool support for considering energy bugs related to location-based apps at the design phase is not widely investigated in the literature. Therefore, in this paper, we aim at investigating the possibility of having tool support for reducing the battery consumption of location-based smartphone applications.
Therefore, we aim to develop a tool, \textit{e}GEN, that consists of DSML and code generator for location-based Android applications.
The contributions of this paper are three-fold:
\begin{enumerate}
    \item A Textual Domain-specific Modeling Language (DSML) for creating battery-aware and self-adaptive energy-saving location sensing-strategies.
    \item A code generator to generate the native battery-aware Java code from self-adaptive energy-saving location sensing-strategies.
    \item Controlled experiment based evaluation of the \textit{e}GEN generated code on five open-source location-based Android applications, where \textit{e}GEN's adaptive code produced a savings of 188mA in battery and 4.35 minute reduction in GPS usage per hour at the expense of 97 meters degrade in accuracy over a distance of 3060 meters. We have done evaluation through \textit{e}GEN and non-\textit{e}GEN versions. 
\end{enumerate}

\textit{e}GEN is designed to consider the impact of different context information such as \textit{battery charging state, battery level, foreground or background app execution, and sensing-interval} on energy-saving. 
Further, the code generator of \textit{e}GEN generates the Java code that could be added to the existing Android repositories to make them battery-aware. 
\textit{e}GEN is developed using \textit{Xtext and Xtend} as an Eclipse plugin. 
The \textit{Xtext} is used to define the language elements, and \textit{Xtend} is used to define the code generator. 
\textit{e}GEN currently supports the Android platform and covers the \textit{battery manager API, fused location provider client, and Android activity life cycle}. 
In future, we aim to support the iOS platform as well. 
The source code of \textit{e}GEN is available on GitHub \footnote{\url{https://github.com/Kowndinya2000/egen}} as an open-source repository for others to use.

The efficacy of \textit{e}GEN has been evaluated through controlled experiments. 
We have selected five open-source Android applications that primarily use the location for its operation. 
We have used DSML of \textit{e}GEN to specify self-adaptive location-sensing for all subject applications. 
The code generated using \textit{e}GEN has been instrumented to the subject application to check the improvements in energy-saving. 

The controlled experiments were conducted on \textit{e}GEN and non-\textit{e}GEN versions of the subject applications.
The Google Battery Historian was used to estimate the GPS active time and battery consumption. 

The experiment results show that \textit{e}GEN provides the ability to specify self-adaptive location-sensing strategies. 
In addition, the code generator can generate Java code that can be instrumented to the existing apps without affecting its existing functionalities. 
The subject applications were instrumented using the generated code and executed on Nokia C3\footnote{\url{https://www.nokia.com/phones/en_in/nokia-c-3?sku=SP01Z01Z2428Y}} smartphone running on Android 10 platform. 
Overall, the \textit{e}GEN version of the subject application shows reduction of 4.35 minutes in GPS active time and 188 mA in battery consumption.
As observed from the results, \textit{e}GEN generated code show reduced battery usage and negligible accuracy degradation, showing initial promise and the need for further research.

The rest of the paper is organized as follows: we provide the necessary background information about location-based apps along with a motivating scenario in Section \ref{sec:backg}. Section \ref{sec:design} presents the complete description of \textit{e}GEN's design, which is then followed by tool evaluation in Section \ref{sec:eval}. Threats to validity and related work are discussed in Sections \ref{sec:threat} and \ref{S:related} respectively. Finally, with pointers to the future work, the paper is concluded in Section \ref{sec:conclusion}.

\section{Background and Motivating Example}
\label{sec:backg}
This section presents the background information about the location-based application, the need for self-adaptive location-sensing, and a motivating example.
\subsection{Location-based Applications}
User location is important context information \cite{schmidt1999there} to provide location-based services to smartphone users. 
Smartphone users widely use Location-Based Services (LBS) \cite{dey2009location} for map navigation, discovering nearest places of interest, activity or mobility tracking, trajectory monitoring, location-based social networking, games, advertisement, and weather forecasting \cite{huang2018location}.
New generation smartphones can handle all types of location-based services with their rich hardware and software capabilities \cite{voicu2019human}. 
The smartphones are equipped with GPS, WiFi \cite{choi2017energy}, Cell ID \cite{ibrahim2012cellsense}, accelerometer, magnetometer, barometer, gyroscope \cite{yang2015mobility}  to position the users' current location. 
Global Positioning System (GPS) \cite{hofmann2012global} is widely used to locate the current location of the user. 
GPS provides high accuracy of user location compared to other location-sensing techniques such as WiFi positioning \cite{choi2017energy} and Cell ID based positioning \cite{ibrahim2012cellsense}. 
One downside of GPS is its abnormal amount of energy consumption for its operation \cite{oshin2012improving}.
Therefore, GPS usage must be reduced \cite{paek2010energy} to extend the smartphone's battery life. 
Unfortunately, GPS usage is a inevitable requirement in map navigation, activity tracking, and trajectory monitoring applications.

The following strategies were found in the literature to address the abnormal battery draining issues of GPS: (1) GPS Alternatives, (2) Movement Detection, (3) Collaborative Strategies, and (4) Adaptive Strategies. \textit{GPS Alternatives} related research efforts use the energy-efficient alternatives such as Cell-ID sequencing matching \cite{paek2011energy}, GSM positioning \cite{ibrahim2012cellsense}, and WiFi-based positioning \cite{choi2017energy, mariakakis2014sail}.
These approaches are affected by poor location accuracy, which is not suitable for continuous location-sensing. 
The research efforts under \textit{Movement Detection} category includes approaches such as scheduling of location updates \cite{kjaergaard2009entracked}, turning off GPS when not  available \cite{paek2010energy}, and postponing GPS updates \cite{cho2015placewalker}. 
These approaches either turn off the GPS updates or delay them by combining inertial sensors like accelerometer, compass, gyroscope, barometer, etc.
In \textit{Collaborative Strategies}, the location coordinates are fetched from the other location-based apps \cite{man2014energy} or neighboring devices \citep{xi2015energy}. 
In \textit{Adaptive Strategies} several research efforts uses energy-accuracy requirements \cite{zhuang2010improving, lin2010energy}, indoor-outdoor detection \citep{capurso2017android}, and context information \cite{kim2016adaptive} to dynamically switch between location sources. 
These adaptive strategies dynamically select a suitable location strategy based on the dynamic accuracy and energy requirements.
Adaptive strategies are better approaches than other approaches as they consider balancing energy and accuracy requirements of location-based services. 

As reported in  \citet{fonseca2019manifesto}, \textit{dynamic adaptability} could be one of the important practices to improve software energy efficiency. 
Therefore, introducing \textit{dynamic adaptability} in existing location-based services would reduce the abnormal battery consumption while satisfying accuracy requirements. 
Nonetheless, developing such self-adaptive behavior is a difficult task for the developers \cite{krupitzer2015survey} as it has to deal with a dynamically changing environment. 
In particular, identifying relevant context information for energy-saving at run-time is not a straightforward task \cite{capurso2018survey}. 
On the other hand, few researchers have proposed generic solutions to address energy-efficiency issues of Android applications \cite{ortiz2019improving, CANETE2020106220} by combining self-adaptivity and battery awareness. 
In this paper, we have investigated the impact of a concept called ``\textit{self-adaptive location-sensing}" by combining \textit{dynamic adaptability} and \textit{battery awareness}. 
The fundamental idea of self-adaptive location-sensing is to enforce energy-saving policies in the following situations: (1) when the battery is discharging, and the battery level is critical, (2) when the app is in the background.


\subsection{A Motivating Example}
\begin{figure}
\centering
\begin{minipage}{.5\textwidth}
    \centering
    \includegraphics[width=\linewidth]{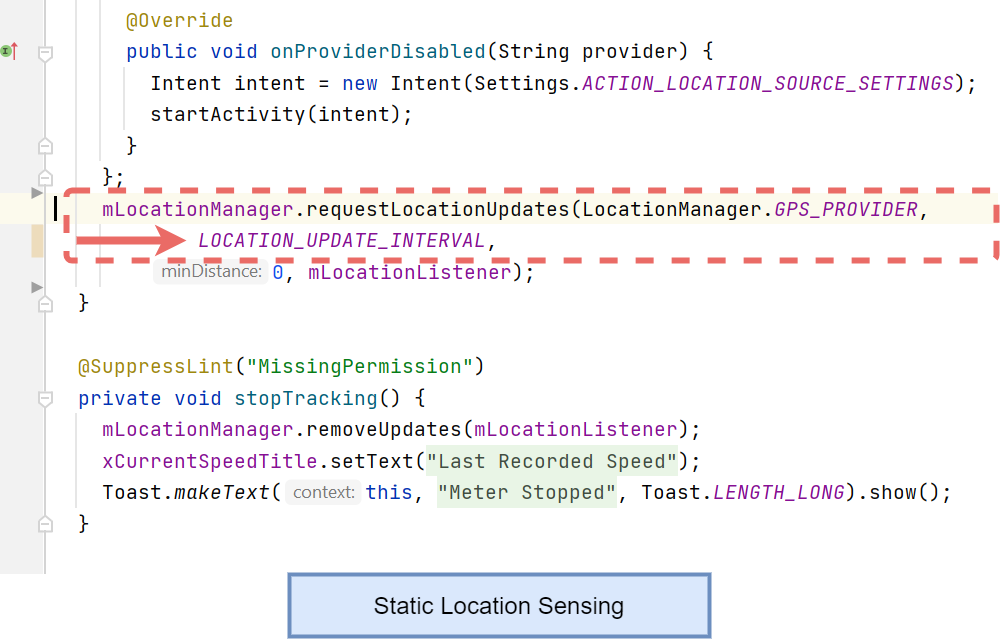}
    \caption{Static location-sensing}
    \label{fig:code_me2}
\end{minipage}%
\begin{minipage}{.5\textwidth}
    \centering
    \includegraphics[width=0.8\linewidth]{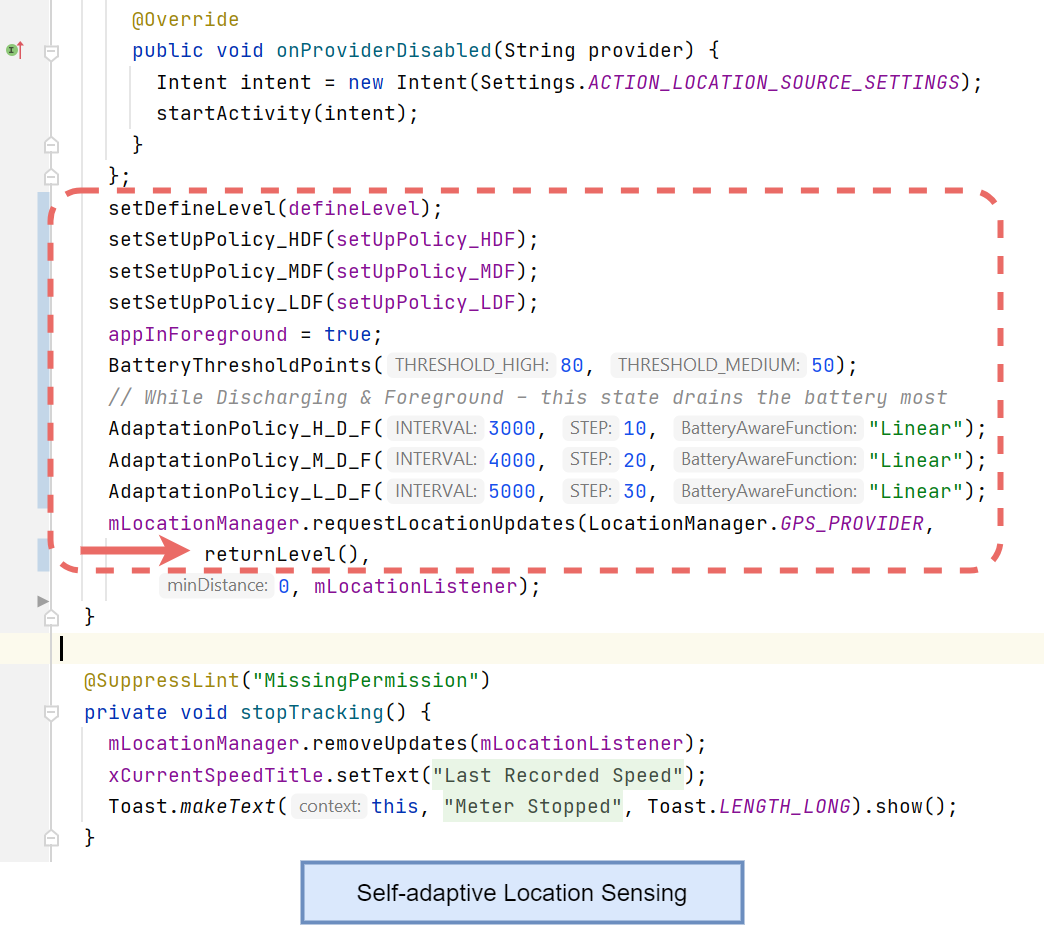}
    \caption{Self-adaptive location-sensing}
    \label{fig:code_me}
\end{minipage}
\end{figure}

\begin{figure}
    \centering
    \includegraphics[width=0.7\linewidth]{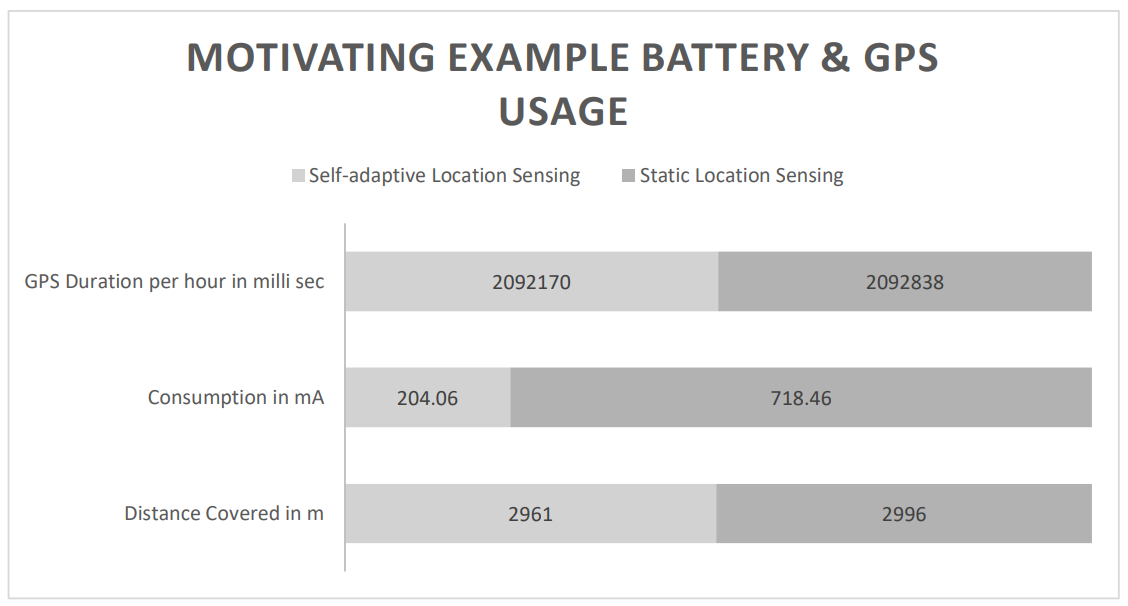}
    \caption{Results of self-adaptive and non-self-adaptive versions}
    \label{fig:energy_me}
\end{figure}

This subsection presents the need for self-adaptive location-sensing through an android application named Speedometer\footnote{\url{https://github.com/iAhmedAwad/Speedometer}}. It calculates the vehicle speed on a real-time basis using GPS. 
This application uses system location service and requests for location updates for a specific sensing interval. 
As highlighted in Figure \ref{fig:code_me}, the Speedometer app uses fixed sensing interval. The sensing interval (time gap between two location calls) plays a significant role in the battery consumption by GPS. The lesser the sensing interval results in higher location accuracy and more battery consumption. In the case of \textit{Speedometer} app, the sensing interval is maintained less and fixed for more accuracy. This usage scenario is suitable when the battery level is high and connected to the charger. The fixed sensing interval might deplete the battery quickly if the battery is discharging. The quick battery discharge problem could be addressed by increasing the sensing interval. However, the location accuracy might degrade if the sensing interval is increased randomly. Hence, the sensing interval must be made adaptive in response to the change in battery level to balance location accuracy and battery consumption requirements. We have experimented with a speedometer app to calculate the GPS active time, battery consumption, and distance measured. As shown in Figure \ref{fig:energy_me}, the original Speedometer app shows 2092838 ms of GPS active time\/hr, 718.46 mA battery consumption, and 2996 m estimated distance. 

To illustrate the efficacy of self-adaptive location-sensing, we instrumented the Speedometer app source code to make it battery-aware. In Figure \ref{fig:code_me2}, the lines of code inside the dashed red box represent the self-adaptive code added to the original version of the Speedometer app. Predominantly, the function $BatteryThresholdPoints$ takes the battery thresholds as $HIGH, MEDIUM, LOW$, depending on the remaining battery level. 
The functions $AdaptationPolicy\_H\_D\_F$ determine the dynamic sensing interval by considering the base sensing interval and decreasing factor.
Here, $H$ refers to the battery state \textit{HIGH}, $D$ refers to battery status \textit{Discharging}, $F$ refers to the application state \textit{Foreground}.
In Figure \ref{fig:code_me2}, the arrow points to the function $returnLevel$ calculates the dynamic location-sensing interval based on the context values. We have also defined other functions $AdaptationPolicy\_M\_D\_F$ and $AdaptationPolicy\_L\_D\_F$ to calculate suitable sensing interval based on the current battery level and app state.   

As shown in Figure \ref{fig:energy_me}, the self-adaptive location-sensing significantly reduces the battery consumption (514.5 mA) compared to the original app with static sensing interval. On the other hand, it is observed that the accuracy degrade only 35 meters which is a consequence of reduced GPS active time per hour. The results show that the self-adaptive location-sensing might help developers reduce significant battery consumption by increasing the sensing interval with lesser degrade in accuracy. We have carefully selected the self-adaptive location-sensing policies as battery and location accuracy requirements are conflicting. Specifically, the impact of battery level drop and sensing interval increasing rate for each battery drop plays a significant role in determining energy-saving adaptation policies. Since this approach involves handling multiple context information such as battery level, battery charging state, and sensing interval increasing factor, it might be wiser to decide the sensing policies before the development phase. Hence, there is a need for a suitable methodology and tool support to analyze application requirements and define appropriate energy-saving location-sensing policies. Therefore, this paper aims to present domain-specific language support to define location-sensing policies before development. In addition, we aim to provide a code generator that the developers may use to introduce battery awareness. 

\section{\textit{e}GEN Design and Development}
\label{sec:design}
With this paper, we would like to address the following questions: 
\begin{tcolorbox}
\begin{itemize}
    \item  What factors influence battery consumption and how to fine tune those factors?
    \item How to make battery-aware decisions during design time?
    \item  How to plug battery-aware adaptation into an existing android application?
\end{itemize}
\end{tcolorbox}
Our attempt towards answering all the above questions is the \textit{e}GEN framework. 
\color{black}
This section presents the overview, grammar definition, and language elements of DSML developed as part of \textit{e}GEN.
\subsection{Overview}
\begin{figure*}[t]
\centerline{\includegraphics[width=0.9\linewidth]{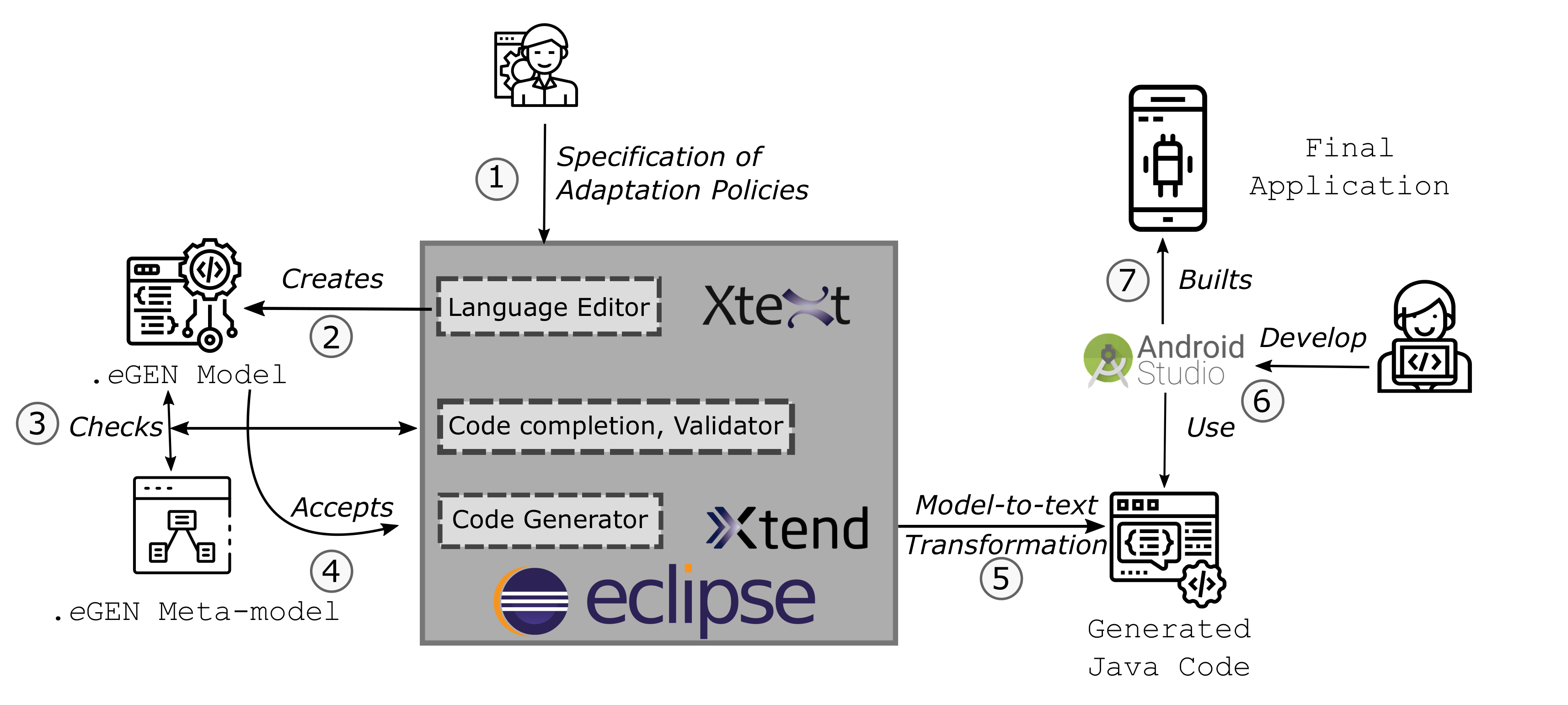}}
\caption{\textit{e}GEN eco-system}
\label{fig:process}
\end{figure*}

The essential idea of energy-saving in location-sensing is to enforce energy-saving policies in the following situations: (1) when the battery is discharging, and the battery level is critical, (2) when the app is in the background. 
Therefore, \textit{e}GEN is designed to assign values for \textit{critical battery level, and sensing-interval} based on the application requirements. 
\textit{e}GEN consists of a domain-specific modeling language and automatic code generator. It has been developed, with the help of \textit{Xtext and Xtend} \cite{behrens2008xtext}.
As shown in Figure \ref{fig:process}, the usage of \textit{e}GEN consists of seven steps:
\newline 
\begin{steps}
    \item The Eclipse editor is used to specify the energy-saving location-sensing policies using the textual domain-specific modeling language.
    \item The editor creates the ${.egen}$ model.
    \item The \textit{validator} module of \textit{Xtext} checks the ${.egen}$ model whether it is following the \textit{e}GEN DSML grammar.
    \item The code generator takes the validated ${.egen}$ model as the input.
    \item It then generates the Java code using model-to-text transformation. 
    \item The generated code can be added to existing Android applications to make it energy-aware.
    \item The updated Android project can be built and installed on the Android device by the user.
\end{steps}

\subsubsection{Domain Model}
As shown in Listing \ref{lis:domain-model}, the domain model contains the entities \\${AdaptationPolicy, Context, Feature}$. 
\begin{lstlisting}[caption=Domain Model for Modeling Location-sensing, captionpos=b, basicstyle=\ttfamily\footnotesize, label=lis:domain-model, emph={AdaptationPolicy, Context, Adaptation},emphstyle={\bfseries}]
entity AdaptationPolicy {
	PolicyID
	Context
	Adaptation
	}
entity Context {
	BatteryState = (Charging | Discharging)
	BatteryLevel = (Low | Medium | High)
	ApplicationState (ForeGround | Background)
	}
	
entity Feature {
	SensingInterval
	DecreasingFactor
	BatteryAwareFunction
	}
\end{lstlisting}
The ${AdaptationPolicy}$ refers to the self-adaptive energy-saving location-sensing strategies. 
It can be defined with the combination of ${Context}$ and ${Feature}$. 
The entity ${Context}$ refers to the contextual situation suitable for enabling energy-saving self-adaptation. 
The context information includes ${BatteryState}$, ${BatteryLevel}$, and ${ApplicationState}$.
The ${BatteryLevel}$ refers to the amount of battery (in \%) left in the smartphone. 
The ${BatteryState}$ refers to the charging or discharging status of the smartphone. 
The ${ApplicationState}$ refers to the background or foreground execution of the application. 
The entity ${Feature}$ includes ${SensingInterval}$, ${DecreasingFactor}$, and ${BatteryAwareFunction}$.

The ${SensingInterval}$ refers to the time difference between two subsequent location-sensing requests. 
The ${DecreasingFactor}$ refers to the numerical value that will be used to calculate the sensing interval for each battery drop in the exponential battery-aware function.
The ${BatteryAware Function}$ refers to the type of change (exponential or linear) in fixing the sensing interval. 
This domain model is used as a basis for defining the grammar of the DSML, which is part of \textit{e}GEN.

\subsection{DSML Grammar}

This subsection describes the \textit{e}GEN grammar along with the structure of the language elements such as ${features}$, ${context}$, and ${adaptation policy}$. 

\subsubsection{Allowed features}
The ${feature}$ element of \textit{e}GEN is used to define the application requirements that affect the battery consumption of smartphone devices.
Especially in location-based services, the features like \textit{location-sensing interval and type of change in sensing interval} play a significant role in deciding the self-adaptive location-sensing strategies.

\begin{lstlisting}[frame=single, numbers=left, basicstyle=\ttfamily\footnotesize,  breaklines=true, caption=Structure of the features values, label=lis:feature, emph={feature,variants},emphstyle={\bfseries}]
Features:
    SensingInterval | Decreasing_Factor | BatteryAwareFunction;
\end{lstlisting}
As shown in Listing \ref{lis:feature}, \textit{e}GEN allows following ${Features}$ for specifying energy-aware requirements:
\begin{itemize}
\item ${SensingInterval}$ refers to the time difference between two subsequent location-sensing requests. 
\item ${DecreasingFactor}$ refers to the numerical value that will be used to calculate the sensing interval for each battery drop in the exponential battery-aware function. 
\item ${BatteryAwareFunction}$ refers to the type of change (exponential or linear) in fixing the sensing interval.
\end{itemize}
\subsubsection{Feature definition}
Each ${Feature}$ can have their own rules for defining the corresponding values as shown in Listing \ref{lis:feature-defn}.
The rules for defining feature values are given below:
\begin{itemize}
	\item The definition of location-sensing interval starts with the keyword "${SensingInterval}$" and can be assigned with an ${integer}$ value (refer $\textit{lines 1-2}$ in listing \ref{lis:feature-defn}). Here, the "${SensingInterval}$" must be assigned in milliseconds. 
	\item The definition of decreasing factor starts with the keyword "${DecreasingFactor}$" and can be assigned with the ${integer}$ value (refer $\textit{lines 3-4}$ in listing \ref{lis:feature-defn}) as decided by the domain analyst. 
	\item The definition of a type of battery-aware function starts with defining the value for keyword "${BatteryAwareFunction}$" and can be assigned with one of the following fixed values: ${linear, exponential}$ (refer $\textit{lines 5-6}$ in listing \ref{lis:feature-defn}).
\end{itemize}
\newpage
\begin{lstlisting}[frame=single, numbers=left, basicstyle=\ttfamily\footnotesize,  breaklines=true, caption=Structure of feature definition, label=lis:feature-defn, emph={vconstraint,validvariant,==,AND,then,isInvalid},emphstyle={\bfseries}]
SensingInterval:
    'SensingInterval' '=' ivalue = MYINT_T;
DecreasingFactor:
    'DecreasingFactor' '=' ivalue=MYINT_T;
BatteryAwareFunction:
    'BatteryAwareFunction' '=' value=('Exponential' | 'Linear');
\end{lstlisting}
\subsubsection{Allowed context}
The ${Context}$ element of \textit{e}GEN is used to define the valid situations to enforce self-adaptive energy-saving policies of smartphone applications. For location-based services, the following context is considered in \textit{e}GEN: \textit{remaining battery percentage, charging state of the device, and state of the application}. 
\begin{lstlisting}[frame=single, numbers=left, basicstyle=\ttfamily\footnotesize, breaklines=true, caption=Structure of the context values, label=lis:context, emph={context,values},emphstyle={\bfseries}]
Context:
    BatteryState | BatteryLevel | AppState | Threshold_Medium | Threshold_High
\end{lstlisting}
As shown in Listing \ref{lis:context}, \textit{e}GEN allows following ${Context}$ for specifying energy-saving situations:
\begin{itemize}
\item ${BatteryState}$ refers to the charging or discharging status of the smartphone. 
\item ${BatteryLevel}$ refers to the remaining battery percentage of the smartphone. Further, the context ${Threshold\_High}$ and ${Threshold\_Medium}$ is used to define the high and medium battery percentage for triggering self-adaptive behavior.
\item ${ApplicationState}$ refers to the background or foreground execution status of the application. 
\end{itemize}
\subsubsection{Context definition}
According to \textit{e}GEN grammar, the definition of each allowed context can have its pre-defined values, and domain analyst defined values.
\begin{lstlisting}[frame=single, numbers=left, basicstyle=\ttfamily\footnotesize,  breaklines=true, caption=Structure of the context constraints, label=lis:context-defn, emph={cconstraint,validcontext,==,AND,then,isInvalid},emphstyle={\bfseries}]
BatteryState:
    'BatteryState' '=' value=('Charging' | 'Discharging');
BatteryLevel:
    'BatteryLevel' '=' value=('High' | 'Medium' | 'Low');
Threshold_High:
    'Threshold_High' '=' ivalue=MYINT_T;
Threshold_Medium:
    'Threshold_Medium' '=' ivalue=MYINT_T;
AppState:
    'AppState' '=' value=('Foreground' | 'Background')
\end{lstlisting}
As shown in Listing \ref{lis:context-defn}, the rules for specifying context values are given below:
\begin{itemize}
	\item The definition of charging state of the device starts with the keyword "${BatteryState}$"  and can have one of the following values: ${Charging}$, ${Discharging}$ (refer $\textit{lines 1-2}$ in listing \ref{lis:context-defn}).
	\item The definition of remaining battery level starts with the keyword "${BatteryLevel}$" and can have any one of the following values: ${High}$, ${Medium}$, ${Low}$ (refer $\textit{lines 3-4}$ in listing \ref{lis:context-defn}). The values for ${High}$ and ${Medium}$ can be assigned with an integer value based on the application requirements (refer $\textit{lines 5-8}$ in listing \ref{lis:context-defn}). The value ${Low}$ will be inferred by the code generator script based on the range given for ${High}$ and ${Medium}$.
	\item The definition of application execution state starts with the keyword "${AppState}$" and can have any one of the following values: ${Foreground}$, ${Background}$ (refer $\textit{lines 9-10}$ in listing \ref{lis:context-defn}).
\end{itemize}

\begin{lstlisting}[frame=single, numbers=left, basicstyle=\ttfamily\footnotesize,  breaklines=true, caption=Structure of the adaptation policy, label=lis:adap-policy, emph={Model, AdaptationPolicy,Condition,Adaptation,=,AND},emphstyle={\bfseries}]
Model:
    eGEN += AdaptationPolicy*;
AdaptationPolicy:
    'AdaptationPolicy' ivalue=MYINT_T '{' 'Condition' '{' Situation1 = Context 'AND' value=(Situation2);
Situation2:
    Block = Situation3 '}' 'then' 'Adaptation' '{' FeatureBlock1 = Features 'AND' value=(FeatureBlock2)  '}' '}';
Situation3:
    Context = Context 'AND' value=(Situation4);
Situation4:
    Context = Context 'AND' value=(Situation5);
Situation5:
    Context = Context 'AND' value=(Context);
FeatureBlock2:
    Feature2 = Features 'AND' value=(Features);
\end{lstlisting}

\subsubsection{Adaptation policy}
According to \textit{e}GEN grammar, the definition of adaptation policy consists of assigning \textit{five} ${contexts}$ and \textit{three} ${features}$. The specification of a self-adaptive location-sensing policy can have one or more entries differentiated with a unique ID.
As shown in Listing \ref{lis:adap-policy}, a single adaptation policy definition contains following parts:
\begin{itemize}
\item starts with the keyword ${AdaptationPolicy}$ to describe an adaptation policy followed by the unique ID of type ${integer}$ (refer $\textit{lines 3-4}$ in listing \ref{lis:adap-policy}). 
\item an opening brace for adaptation policy definition
\item a keyword ${Condition}$ to describe the allowed context changes that trigger the self-adaptation
\item an opening brace for context block definition
\item five condition definition, each consists of context assigned with allowed values. The description of context values is given in Listing \ref{lis:context-defn}. Here, the context definition can be in any order. However, the repetition of context information is not allowed inside the same context block.
\item multiple valid contexts can be separated by the keyword ${Condition}$. 
\item a closing brace for context block
\item a keyword ${Adaptation}$ to describe the corresponding set of features to be executed at run-time for the contextual changes described with the keyword ${Condition}$. 
\item an opening brace for a feature block
\item three feature definition, each consists of features assigned with allowed values. The description of feature definition is given in Listing  \ref{lis:feature-defn}. Here, the feature definition can be in any order. However, the repetition of feature information is not allowed inside the same feature block.
\item multiple adaptations can be separated by the keyword ${Condition}$. 
\item a closing brace for a feature block
\item Finally, a closing brace for adaptation policy
\end{itemize}

\begin{figure}
    \centering
    \includegraphics[width=0.8\linewidth]{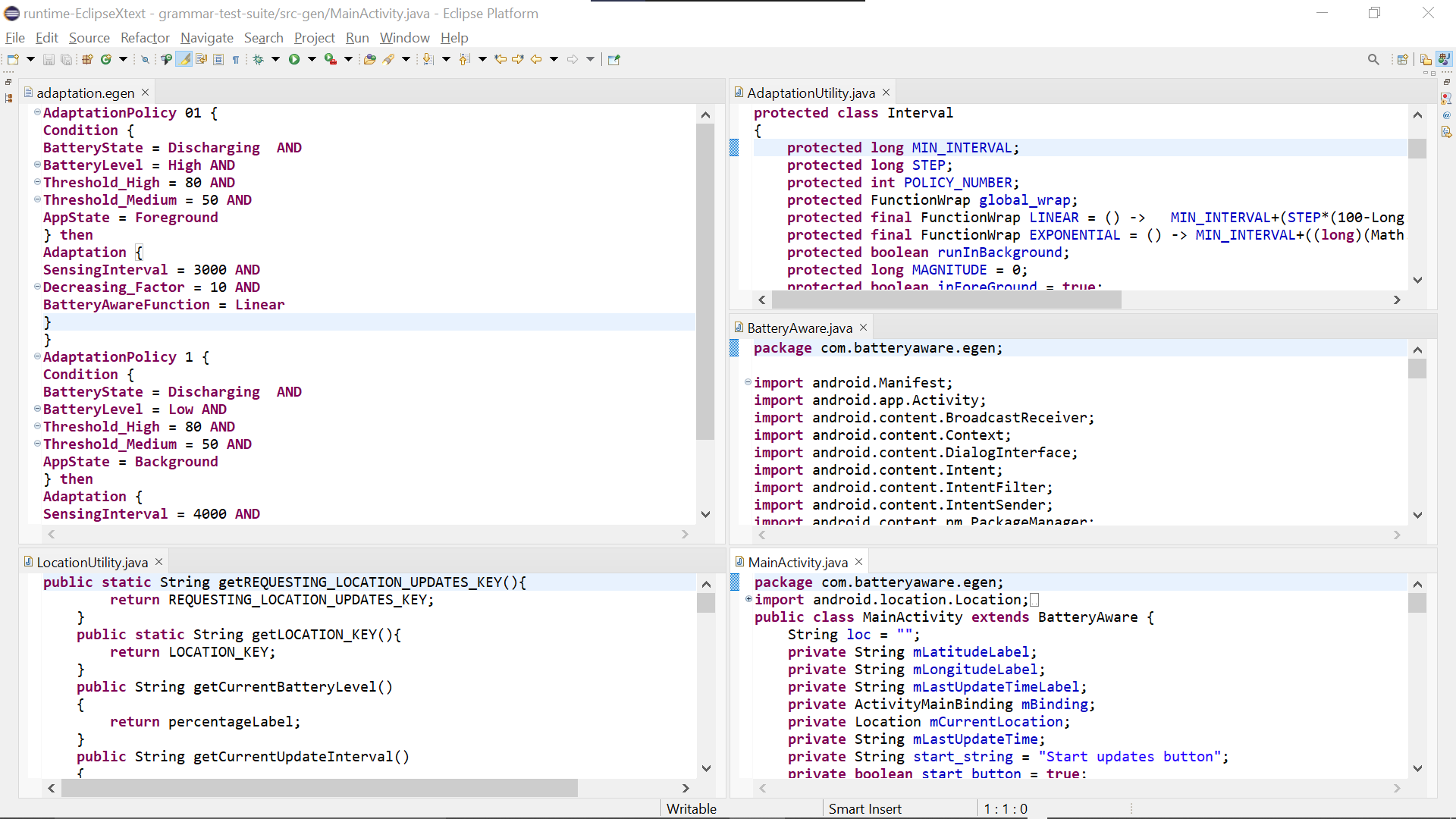}
    \caption{A sample .egen model and the generated code}
    \label{fig:code_generator}
\end{figure}
\subsection{Code Generator}
The code generator uses the following Android APIs to achieve self-adaptation: \textit{Battery Manager API, Fused Location API, and Android Activity Life Cycle}.
The code generator is defined by mapping each element in the \textit{e}GEN DSML to a corresponding Android library.
The $\textit{BatteryManager}$ class is used to fetch the $\textit{BatteryState, BatteryLevel}$ of the Android device through variables ${BATTERY\_PROPERTY\_CAPACITY}$ , and ${BATTERY\_PROPERTY\_STATUS}$. 
The application's status is identified using the Android activity lifecycle $\textit{onStart()}$, $\textit{onResume(), onPause()}$. 

As shown in the Figure \ref{fig:code_generator} the code generator creates Java files ($\textit{MainActivity.Java}$, $\textit{LocationUtility.Java}$, $\textit{BatteryAware.Java}$, and $\textit{AdaptationUtility.java}$) that contains the artifacts for the self-adaptive location-sensing.
$\textit{BatteryAware.java}$ is the file that does the adaptive location-sensing activity. 
MainActivity extends the $\textit{BatteryAware}$ activity and fetches the location coordinates from the function $\textit{onLocationUpdate()}$ defined in the batteryAware class. 
The application developers can modify $\textit{MainActivity.java}$ to write their business logic. The file $\textit{AdaptationUtility.java}$ contains the code that alters the sensing-interval based on the context provided. The battery state defined in the \textit{egen} model is verified against the charging status obtained from the $\textit{BatteryManager}$ API. $\textit{LocationUtility.java}$ contains the code that does the location fetching activity as per the sensing-interval interval captured in the $\textit{AdaptationUtility.java}$. The generated code can be appended to the existing Android projects to make their app self-adaptive for location-sensing.  

\section{Evaluation of \textit{e}GEN}
\label{sec:eval}
In this section, we evaluate the effectiveness of \textit{e}GEN at reducing the battery consumption of open-source location-based Android applications. The code generated by \textit{e}GEN has been instrumented to the subject applications to show its efficacy. We call our evaluation controlled experiments as the test smartphone's battery is only subjected to the application being monitored and some mandatory operating internal services which will use battery all the time (eg. screen, operating system). All other apps or services that can be uninstalled or stopped have been stopped. This experimental setup enabled us to avoid skewed results and lead us to better investigate the effects of using \textit{e}GEN's adaptive code on battery and GPS consumption. In literature, several studies have used controlled experiments \cite{morales2018earmo, banerjee2018energypatch, pereira2020spelling} to evaluate the effectiveness of tool support. Hence, we have designed a controlled experiments based empirical study to evaluate the effectiveness of \textit{e}GEN. 
\color{black}
Hence, we have designed a controlled experiments based empirical study to evaluate the effectiveness of \textit{e}GEN. 
\subsection{Study Design}
The study was conducted to answer the following research questions:
\begin{itemize}
    \item \textbf{RQ1}: Does code generated by \textit{e}GEN reduce GPS usage?
    \item \textbf{RQ2}: How much battery consumption is reduced by code generated from \textit{e}GEN?
    \item \textbf{RQ3}: To what extent code generated by \textit{e}GEN degrade location accuracy?
\end{itemize}
As the main objective of this study is to analyze the effectiveness of \textit{e}GEN on balancing energy-accuracy requirements, the relevant independent and dependant variables are selected as follows:
\begin{itemize} 
\item \textbf{Independent variable}: The \textit{Battery drain} has been selected as a primary independent variable as location-sensing reduces the battery over-time based on the GPS usage. 
\item \textbf{Dependent variables}: In this study, we have selected the variables, \textit{sensing interval, GPS usage, battery drop rate, and distance covered} as dependent variables. These variables are directly affected by every drop in battery percentage as specified in the adaptation policy. 
\end{itemize}
\subsubsection{Subject Application Selection}
We have used F-Droid\footnote{\url{https://www.f-droid.org/}} and Google Play Store\footnote{\url{https://play.google.com/store}} to find out the subject applications. Specifically, we have searched for relevant apps with keywords such as \textit{gps logging, distance measurement, path tracker, location service, speed meter, location share, live location tracking, etc}. Finally, we have selected the android applications that primarily rely on GPS for their operation. 
We have considered apps if it is open source and the source code is available on code sharing platforms like GitHub\footnote{\url{https://github.com/}}. 
\begin{table}[t]
\caption{List of Subject Applications}
\label{tab:sub-apps-list}
\scriptsize
\begin{tabular}{|p{1em}|p{10em}|p{6em}|p{6em}|p{25em}|}
\hline
\textbf{S.No} & \textbf{App Name}       & \textbf{Play Store} & \textbf{F-Droid} & \textbf{GitHub URL}                                   \\ \hline
1             & GPS Logger & Available                  & Not Available    & \url{https://github.com/BasicAirData/GPSLogger}             \\ \hline
2             & OSM Tracker             & Available                  & Not Available    & \url{https://github.com/labexp/osmtracker-android}          \\ \hline
3             & Runner Up                & Available                  & Available        & \url{https://github.com/jonasoreland/runnerup}              \\ \hline
4             & OpenTracks              & Available                  & Available        & \url{https://github.com/OpenTracksApp/OpenTracks} \\ \hline
5             & KinetiE- Speedometer     & Available                  & Not Available    & \url{https://github.com/xyz-relativity/KinetiE-Speedometer} \\ \hline

\end{tabular}
\end{table}

\begin{table}[t]
\caption{GitHub repository profile of Subject Applications }
\label{tab:sub-apps-stats}
\begin{tabular}{|c|c|c|c|c|}
\hline
\textbf{S.No} & \textbf{App Name}       & \textbf{\# Contributors}    & \textbf{\# Stars}   & \textbf{\# Forks}                                      \\ \hline
1             & GPSLogger               & 6                          & 238                      & 92                                                    \\ \hline
2             & OSMTracker              & 25                         & 432                      & 220           \\ \hline
3             & RunnerUp                & 47                         & 601                      & 260               \\ \hline
4             & OpenTracks              & 60                         & 380                      & 75            \\ \hline
5             & KinetiE-Speedometer     & 1                          & 0                        & 0  \\ \hline
\end{tabular}
\end{table}
The following inclusion criteria were applied to filter the relevant subject applications:  
\begin{enumerate}
    \item If the app is written in Native Android code
    \item If GPS is primarily used for location sensing
    \item If the source code repository published with an open-source license 
    \item If the recent most commit is published less than two years
    \item If the repository is well documented
\end{enumerate}
We found 11 candidate applications by applying the above-mentioned inclusion criteria. Further, the following exclusion criteria were applied on the 11 candidate applications:
\begin{enumerate}
    \item If the app's android support plugin is incompatible with our IntelliJ IDEA version. 
    \item If the app has missing dependencies or other build errors 
    \item If it is not compatible with recent versions of Android (such as 10.0 or 9.0)
    \item If the app activity crashes while installing and using the Android application  
\end{enumerate}
We applied the exclusion criteria as mentioned above and removed the apps that satisfy at least one criteria. Finally, we found \textit{five} subject applications that are suitable for instrumentation and conducting controlled experiments. Table \ref{tab:sub-apps-list}, shows the list of subject applications and their availability on Google Play Store, F-Droid, and GitHub. Table \ref{tab:sub-apps-stats} shows the demographic data gathered from the source code repository like GitHub. Each of the subject applications except KinetiE-Speedometer have at least 6 contributors, 238 stars, and 75 forks. We have chosen KinetiE-Speedometer to verify the usefulness of \textit{e}GEN generated for applications that do not rely on other position sensing sensors. 

\subsubsection{Investigation of Subject Applications and Writing Adaptation Polices}
\begin{figure}
\centering
\begin{minipage}{.4\textwidth}
    \centering
\begin{lstlisting}[frame=single, numbers=left, basicstyle=\ttfamily\scriptsize,  breaklines=true, caption=An excerpt of the instrumented Adaptation Policy, label=lis:ins_adap_policy]
AdaptationPolicy 01 {
    Condition {
        BatteryState = Discharging  AND
        BatteryLevel =  High AND
        Threshold_High = 80 AND
        Threshold_Medium = 50 AND
        AppState = Foreground 
    } then
    Adaptation {
        SensingInterval = 3000 AND
        Decreasing_Factor = 10 AND
        BatteryAwareFunction = Linear 
    }
}
\end{lstlisting}
\end{minipage}%
\begin{minipage}{.6\textwidth}
    \centering
    \includegraphics[width=0.9\linewidth]{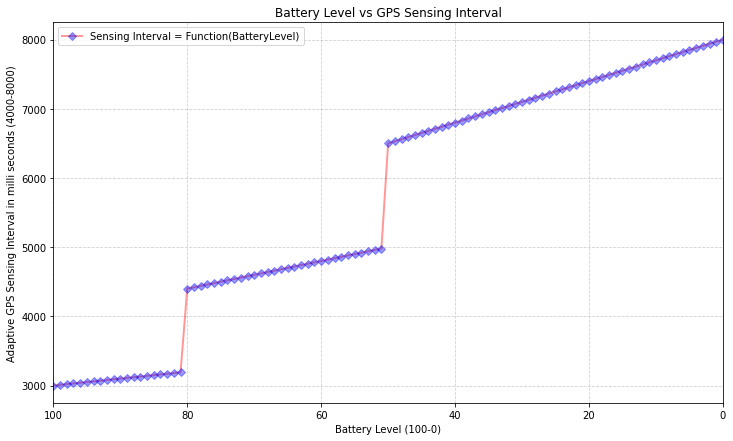}
    \caption{Change in Sensing interval}
    \label{fig:level_vs_gps}
\end{minipage}
\end{figure}

The first and second authors have investigated the subject application's source code to identify the modules responsible for location listeners. We have looked for the location API, sensing interval, and support for battery awareness. 
The subject applications used static sensing intervals and became a candidate for introducing self-adaptive location sensing. 
Apart from other utility files, each of the subject application contained location listeners and files detailing track information to the user. These were the files involved in code instrumentation; while apps used different location manager APIs, the relaxation time between two location updates was found to be static and can be modified independently without disturbing the location fetching process. Hence, the adaptation code being added to subject applications only derives GPS sensing interval based on battery status and is independent of location manager API used in the subject application. Applications such as RunnerUp, GPSLogger provision users to set the GPS sensing interval based on their preference, however the sensing interval adjusted to user's preference is also static; since \textit{e}GEN version of the app should contain adaptive code and that of non-\textit{e}GEN a fixed value, we have disabled the user control on changing the sensing interval in the app and code has been instrumented accordingly. 
\color{black}
As shown in Listing \ref{lis:ins_adap_policy} (partial), we have defined three battery discharging situations to update the location sensing interval. The full adaptation policies can be found in the readme\footnote{\url{https://github.com/Kowndinya2000/egen\#exact-adaptation-policy-instrumented-for-subject-applications}}. The first situation is when the battery level is \textit{High} (80 and above), the second situation is when the battery level is \textit{Medium} (between 50 to 80), and the third situation is when the battery level is \textit{Low} (below 50). 
We have defined an adaptation policy that assigns sensing intervals of 3 seconds, 4 seconds, and 5 seconds respectively for \textit{High, Medium, and Low} battery level when the battery is discharging. In addition, the decreasing factor for High is set to 10 while 20 and 30 are set for \textit{Medium and Low} battery level, respectively. This will ensure that the sensing interval will increase based on the decreasing factor for each battery drop. As shown in Figure \ref{fig:level_vs_gps}, the sensing interval is increasing at a slower pace along with each battery drop when the smartphone's battery level is more than 80\%. On the other hand, the sensing interval is increasing at a comparatively faster pace when the battery level is \textit{Medium and Low}. Especially, the sensing interval is increasing at a faster rate when the device battery drops to \textit{Low} to reduce the number of GPS calls.
\subsubsection{Instrumentation of Subject Applications}
In this step, we aim at instrumenting the generated code to the subject applications for battery awareness. For this purpose, we downloaded the source code of each subject application from GitHub. We looked for variables that hold the sensing interval, functions that handle location updates, and activities that address map features. We have extracted the generated code from \textit{BatteryAware.java} and instrumented it in the subject application. 
Overall, the instrumentation step consists of following activities: 
\begin{itemize}
    \item \textit{Instrumenting for battery-aware location-sensing}: The subject applications are instrumented with the generated code such that it will dynamically change the location sensing interval based on the current battery level and charging status.
    We found that all the subject applications had static sensing intervals. Therefore, we aim to make it dynamically adaptive in this step. The subject applications are instrumented with the generated code for dynamically changing the location sensing interval based on the current battery level and charging status. The instrumentation initialized the adaptation policies and made a call to the functions described in the \textit{Adaptation Activity} class. Apps such as \textit{GPSLogger, OSMTracker} provided users the option to decide the minimum distance between two location updates and minimum location accuracy. In such situations, we disabled the original code and replaced it with the generated code to make it consistent with other subject applications.
    The subject applications had static GPS location update interval. We brought Battery level and status in context to determine the sensing interval on demand for each location update. This is achieved by initializing the adaptation policies and calling the functions described in the Adaptation Activity class. 
    \item \textit{Instrumenting for estimating distance covered}: This phase aims to instrument the app to fetch the location coordinates and calculate the distance covered in meters. As pointed earlier, in subject applications such as \textit{GPSLogger, OSMTracker}, the distance measurement is given by default. However, other subject applications cannot measure distance and create a need for measuring distance. Hence, we have added functions to collect latitude and longitude information from respective java classes to measure the distance covered. Finally, the measured distance (meters) is annotated on the map view of the subject applications. While some subject applications have distance measurement given by default, others do not measure distance travelled. The activity that displays the user track/location information might not calculate the location related events and use other java classes that do GPS location sensing as function calls. Hence, we independently added additional functions that collect latitude and longitude information from respective java classes and we annotated the map view with the distance travelled in meters.  
    \item \textit{Rebuilding the application and install on the test device}: After the instrumentation phase, the project has been cleaned and re-built for conducting experiments. We created two versions of executables for each subject application, namely, \textit{e}GEN version and non-\textit{e}GEN version. 
    The \textit{e}GEN version represents the executable built from the instrumented project with battery-aware code. On the other hand, the non-\textit{e}GEN version represents the executable built from the source code downloaded from GitHub. After we instrumented the app,   we cleaned the project and re-built it. We then installed the modified app on the test device and reviewed the app by checking all the UI components and we looked for any activity crashes before doing the actual trail.   
\end{itemize}
Finally, the executables were installed on the test device, and the UI components were verified to ensure that the instrumentation did not affect the app's behavior. We also verified that the apps are not crashing before doing the actual experiments. 
\subsubsection{Experiment Protocol}

We have selected Nokia C3 as a test device for conducting the experiments.
Nokia C3 comes with a 5.99 inch display, and its hardware packs 3GB RAM and 32GB in-built storage with a 3040 mAH battery. 
This smartphone runs on Android 10 and comes with a cleaner version of Android, which helps us disable all the apps to create an isolated environment for conducting controlled experiments. 
Each subject application has been given network/location-related permissions during the trial.  
\begin{figure}
    \centering
    \includegraphics[width=0.9\linewidth]{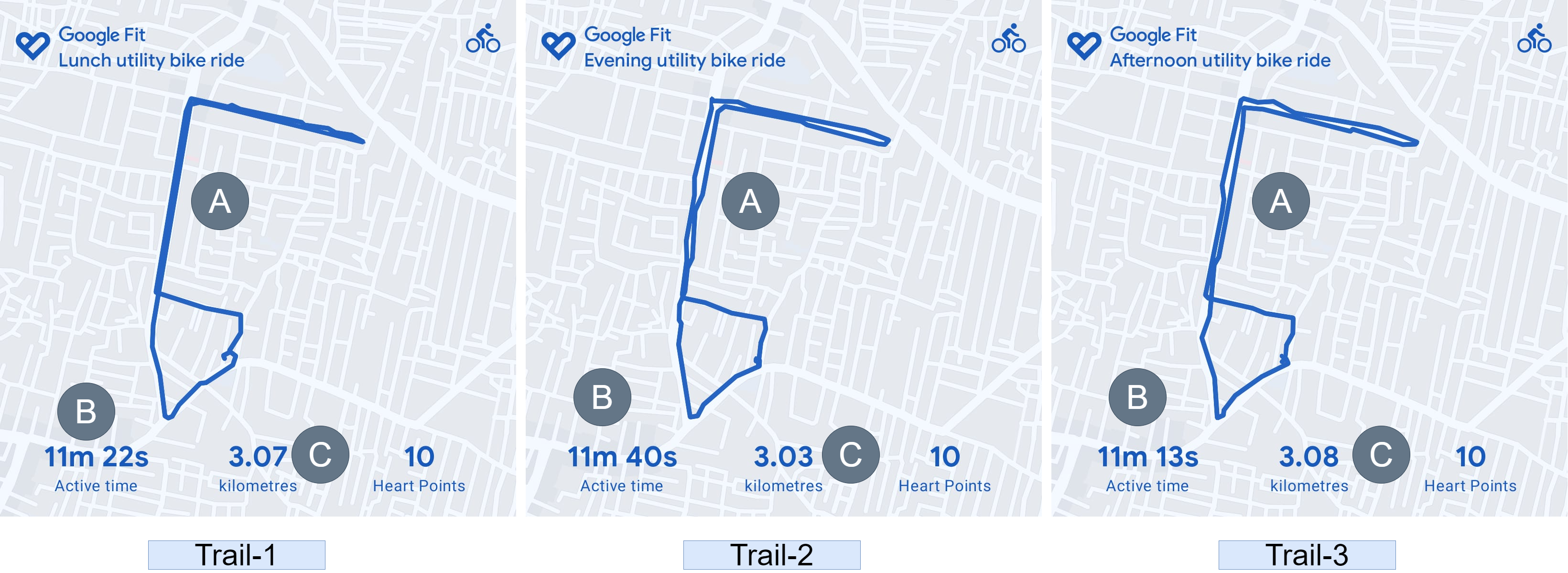}
    \caption{Distance Measured by Google Fit}
    \label{fig:google_fit}
\end{figure}
Initially, we had conducted cycling from the specified source to destination using Google Fit\footnote{\url{https://www.google.com/fit/}} to measure the total distance. The three trails of distance measurement with Google Fit is depicted in Figure \ref{fig:google_fit}. Figure \ref{fig:google_fit}[A] shows the direction of travel from source to source(destination), Figure \ref{fig:google_fit}[B] and [C] show the duration of the trail and distance covered. Three Google fit trials with the same source and the destination have been conducted and found an average distance of 3060 meters. The average battery drop percentage was between 3\% to 4\%. The distance measured with Google Fit (3060 m) is considered a reference distance and used to analyze the location accuracy of subject applications.   

We have conducted three trails for both \textit{e}GEN and non-\textit{e}GEN versions of the five subject applications, thus conducting 30 trails each of length 3 kilometers (approx.). 
For each trial, we followed steps mentioned below:

\begin{itemize}
    \item Installing the instrumented App on testing device
    \item Resetting the bug report using Android Device Bridge (adb)
    \item Cycling from the defined source to destination 
    \item Exporting the bug report after reaching the destination
    \item Uploading the bug report in Google Battery Historian to Measure the percentage of battery drop
    \item Record the distance covered in Meters
\end{itemize}
On completing each controlled trial, we use \textit{adb} to generate the android test device's bug report in .zip format. The generated bug report has been uploaded to the Battery historian for further analysis.
\subsection{Experimental Results}

This subsection presents the experimental results and answers to research questions.
\subsubsection*{\textbf{Answering RQ1}: Does code generated by \textit{e}GEN reduce GPS usage?}
The purpose of this research question is to check whether the code generated by \textit{e}GEN reduces the ${GPS Active Time Per Hour}$ compared to the non-\textit{e}GEN version of the subject application. For this purpose, the Google battery historian tool was used to fetch the ${GPS Active Time Per Hour}$ for each subject application. The reduction in ${GPS Active Time Per Hour}$ is calculated as the time difference between the GPS active time by non-\textit{e}GEN version and \textit{e}GEN version of each subject application. A positive value indicates that \textit{e}GEN version has managed to bring down the GPS active duration per hour. 
As shown in Table \ref{tab:gps_saved_rq1}, the experimental results show that \textit{e}GEN generated code reduces GPS active time by a minimum of 58 seconds and a maximum of 6m27s in an hour.
Specifically, the highest reduction in GPS active time was produced by the instrumented version of \textit{GPSLogger} that reduced the GPS active time by 6-7 minutes. 
On the other hand, the instrumented version of \textit{Open Tracks} produced the lowest reduction in GPS active time of 58 seconds. 
Overall, the experimental results show that \textit{e}GEN reduced the GPS active time by 4.35 minutes per hour on average.
\begin{tcolorbox}
We observed a significant reduction in \textit{GPS usage} of 4.35 minutes by increasing the ${SensingInterval}$ with a suitable ${Decreasing\_Factor}$ for each battery drop. The results show that adopting dynamic location-sensing intervals might help developers reduce the GPS usage.
\end{tcolorbox}
\begin{table}[]
\caption{GPS Active Time Reduction of \textit{e}GEN over Non-\textit{e}GEN}
\begin{tabular}{|c|c|c|c|c|c|}
\hline
\multirow{2}{*}{\textbf{S.No}} & \multirow{2}{*}{\textbf{Subject App}} & \multicolumn{2}{c|}{\textbf{GPS Active Time per hour}} & \multirow{2}{*}{\textbf{\begin{tabular}[c]{@{}c@{}}Reduction in GPS Active\\  Time per hour\end{tabular}}}  \\ \cline{3-4}
                               &                                       & \textbf{Non-\textit{e}GEN}            & \textbf{\textit{e}GEN}           &                                                                                                                                                                                                                                  \\ \hline
1                              & GPSLogger                             & 50m40s                       & 44m13s                  & 6m27s                                                                                                                                                                                                                        \\ \hline
2                              & RunnerUp                              & 42m38s                       & 43m36s                  & 6m2s                                                                                                                                                                                                                         \\ \hline
3                              & KinetiE-Speedometer                   & 46m28s                       & 40m27s                  & 6m1s                                                                                                                                                                                                                          \\ \hline
4                              & OSMTracker                            & 48m52s                       & 46m37s                  & 2m15s                                                                                                                                                                                                                          \\ \hline
5                              & Open Tracks                           & 46m7s                        & 45m9s                   & 58s                                                                                                                                                                                                                          \\ \hline

\end{tabular}
\label{tab:gps_saved_rq1}
\end{table}
\subsubsection*{\textbf{Answering RQ2}: How much battery consumption is reduced by code generated from \textit{e}GEN?}
The purpose of this research question is to identify the amount of battery saved by the instrumentation adaptation policies and \textit{e}GEN generated code. We had conducted the trials when the battery level was either low or transitioning from medium to low or high to medium to cover all the possible battery-critical situations. The Google battery historian was used to estimate the battery consumed by each subjection application. It displays battery usage in a chart by taking the bug report as input. We have estimated the battery consumption in mA by filtering the subject application in the battery usage chart with the help of the subject application's package name. The battery consumed by non-\textit{e}GEN and \textit{e}GEN version was compared, and the difference is reported in Table \ref{tab:battery_saved_rq2}. The positive value of \textit{Energy Savings in mA} implies that the significant amount of battery consumption is reduced by \textit{e}GEN version. Overall, \textit{e}GEN generated code was able to reduce battery consumption in five subject applications. 
The results from subject applications show that the \textit{e}GEN version shows energy savings ranged from 89 mA (approx) to 269 mA (approx).
The calculated mean reduction in battery consumption is 188 mA 
As shown in Table \ref{tab:battery_saved_rq2}, the code generated by \textit{e}GEN can bring a maximum of 268.76 mA (approx.) battery savings in \textit{OpenTracks}.

\vspace{1em}
\begin{tcolorbox}
Overall, the instrumented code shows considerable battery savings for the trails that last for 12-14 minutes.
We believe that the \textit{e}GEN generated code might show significant battery savings when the apps are used for a longer time in real-world scenarios.
Hence, writing a battery-aware code might help developers to reduce unwanted battery consumption in location-based Android applications.
\end{tcolorbox}

\begin{table}[]
\caption{Battery Saved by \textit{e}GEN over Non-\textit{e}GEN}

\begin{tabular}{|c|c|c|c|c|}
\hline
\multirow{2}{*}{\textbf{S.No}} & \multirow{2}{*}{\textbf{Subject App}} & \multicolumn{2}{c|}{\textbf{\begin{tabular}[c]{@{}c@{}}Battery Consumption\\ in mA\end{tabular}}} & \multirow{2}{*}{\textbf{\begin{tabular}[c]{@{}c@{}}Energy Savings\\  in mA\end{tabular}}} \\ \cline{3-4}
                               &                                       & \textbf{\textit{e}GEN}                                 & \textbf{Non-\textit{e}GEN}                                 &                                                                                                                                                                                                           \\ \hline
1                              & OpenTracks                            & 541.03                                       & 809.79                                            & 268.76                                                                                                                                                                                                    \\ \hline
2                              & KinetiE-Speedometer                   & 924.77                                      & 1173.78                                           & 249.01                                                                                                                                                                                                   \\ \hline
3                              & RunnerUp                              & 798.77                                       & 983.49                                           & 184.72                                                                                                                                                                                                 \\ \hline
4                              & OSMTracker                            & 506.37                                      & 656.31                                            & 149.94                                                                                                                                                                                                \\ \hline
5                              & GPSLogger                             & 842.92                                       & 931.80                                           & 88.88                                                                                                                                                                                                   \\ \hline
\end{tabular}
\label{tab:battery_saved_rq2}
\end{table}
\subsubsection*{\textbf{Answering RQ3}: To what extent code generated by \textit{e}GEN degrade location accuracy?}
The purpose of this research question is to find out the extent to which location accuracy is degraded at the cost of reducing battery consumption. We have considered the distance measured by Google Fit as the benchmark to measure the deviation in location accuracy of subject applications. Initially, we have conducted three trials from the same source to destination using Google Fit, and the averaged distance of 3.06 km has been considered for the comparison. 
We used Haversine formula\footnote{\url{https://www.movable-type.co.uk/scripts/latlong.html}} for calculating the distance by collecting the coordinates of the previous and current locations and cumulatively adding the distance between every two locations for the entire trail. 

In Table \ref{tab:rq3}, the distance covered by \textit{e}GEN and non-\textit{e}GEN version of the subject applications is presented along with the error in location accuracy compared to the distance measured by Google Fit.
In addition, the degraded location accuracy is reported in Table \ref{tab:rq3}.
The degrade is calculated by subtracting the location error by non-\textit{e}GEN version and \textit{e}GEN. 
In Table \ref{tab:rq3}, the negative degrade values mean less deviation, and positive values signify more deviation in location accuracy. 
Overall, \textit{e}GEN version interestingly reported a thin margin of degradation in accuracy when compared to the non-\textit{e}GEN version of the subject application. 
Distinctly, the application \textit{KinetiE-Speedometer} brought in a more accurate distance measurement of 80 meters while reducing the battery consumption.  
Similarly, OSMTracker showed equally reliable accuracy in distance measurement when compared to that of Non-\textit{e}GEN while results in battery savings of 149.94 mA.
The apps such as \textit{Open Tracks, RunnerUp, and GPSLogger} showed degraded accuracy about 7, 10, and 54 meters, respectively. However, this degraded accuracy comes with significant battery savings, as reported in Answer to RQ2.

\begin{tcolorbox}
Overall, the \textit{e}GEN versions of the subject applications resulted in an average distance measurement of 2968 meters which is 92 meters lesser than the distance measured by Google Fit. As discussed in RQ2, the average battery consumption reduction of 188 mA might make the 97 meters degrade inaccuracy negligible. 
Hence, writing a suitable self-adaptive location-sensing policy would help to balance battery consumption and other conflicting requirements such as location accuracy.
\end{tcolorbox}
\begin{table}[]
\caption{Distance covered by \textit{e}GEN and Non-\textit{e}GEN}
\begin{tabular}{|c|c|c|c|c|c|c|}
\hline
\multirow{2}{*}{\textbf{S.No}} & \multirow{2}{*}{\textbf{Subject App}} & \multicolumn{2}{c|}{\textbf{\begin{tabular}[c]{@{}c@{}}Distance\\ Covered in km\end{tabular}}} & \multicolumn{2}{c|}{\textbf{\begin{tabular}[c]{@{}c@{}}Error in \\ Location Accuracy \end{tabular}}} & \textbf{\begin{tabular}[c]{@{}c@{}}Degrade in \\ Location Accuracy \end{tabular}} \\ \cline{3-7} 
                               &                                       & \textbf{Non-\textit{e}GEN}                                & \textbf{\textit{e}GEN}                               & \textbf{Non-\textit{e}GEN}                                                       & \textbf{\textit{e}GEN}                                                      &                                                                                                   \\ \hline
1                              & KinetiE-Speedometer                   & 2.79                                             & 2.87                                        & 280                                                                     & 200                                                                & -80                                                                                                 \\ \hline
2                              & OSMTracker                            & 2.990                                            & 2.990                                       & 60                                                                      & 60                                                                & 0                                                                                               \\ \hline
4                              & Open Tracks                          & 2983                    & 2976                                         & 77                                                                      & 84                                                                 & 7                                                                                               \\ \hline

5                              & RunnerUp                              & 2.99                                             & 2.98                                         & 80                                                                      & 90                                                                 & 10                                                                                               \\ \hline
6                              & GPSLogger                             & 3.05                                             & 2.996                                        & 20                                                                      & 74                                                                 & 54                                                                                               \\ \hline

7                              & Google Fit (Benchmark Application)    & \multicolumn{5}{c|}{3.06 km}                                                                                                                                                                                                                                                                                                                      \\ \hline
\end{tabular}
\label{tab:rq3}
\end{table}

\subsection{Implications}
We believe that the experimental results lay a promising foundation for adopting the energy-saving self-adaptive location-sensing policies in open source applications in the future for balancing battery and location accuracy requirements. This subsection presents some of our findings that may further assist domain analysts, app developers, API developers, and researchers in handling self-adaptive location-sensing. 

\subsubsection*{Domain Analysts and Domain Experts} This study shows the importance of considering energy-aware requirements in the early stages of software development. The self-adaptive location-sensing framework presented in this approach can balance battery consumption and location accuracy based on the remaining battery level. Hence, we suggest domain analysts introduce self-adaptive behavior to manage conflicting requirements like battery consumption and location accuracy. Therefore, in the future, the domain analyst and domain experts may consider battery-aware self-adaptive solutions for other application domains. Significantly, the domain analyst may develop domain-specific languages and code generators to address application domain-specific energy bugs and corresponding energy-saving solutions for domains like games, social networks, and other resource-intensive applications. 

\subsubsection*{App Developers and API Developers} As shown in the results, the self-adaptive location sensing with battery-awareness significantly impacts battery consumption and GPS active time. Hence, we suggest developers write battery-aware code from the initial development itself to handle energy-hungry components. The presented approach is about adding battery awareness in the application developer-written code. In addition, we suggest API developers add support for battery awareness in the library to reduce amount development efforts of application developers. For instance, prospective API developers may modify the existing location APIs or create a new API with battery-aware self-adaptive location-sensing. In addition, API developers from other application domains also introduce battery-awareness in the capabilities provided by their library. 

\subsubsection*{Researchers} 
The DSML and code generator presented in this paper shows initial promise for considering domain-specific energy bugs in the early stages. Hence, we suggest potential researchers identify and catalog the domain-specific energy bugs and their possible solutions to help developers make energy-saving decisions during the development phase. Further, researchers may provide suitable domain-specific tool support to handle the most commonly occurring domain-specific energy bugs. 

\section{Threats to Validity}
\label{sec:threat}
In this section, the potential threats to the validity of the presented case study are discussed.
The guidelines given by \citet{runeson2009guidelines} are followed to categorize and discuss the threats.
\subsection*{Construct Validity} Construct validity refers to the degree to which the analysis measures what we aim to study. 
In this study, we aim to measure the \textit{GPS usage, Battery Consumption, and Location Accuracy}. 
There is a high possibility of error in measuring these variables as we do not have widely accepted tools. 
In literature, several studies use hardware tools to measure energy consumption and use older Android phones \cite{liu2014greendroid, morales2018earmo, banerjee2018energypatch}, which might not suit recent Android devices. 
To address this issue, we have decided to use the recent tool developed by Google developers, namely Google Battery Historian. 
The \textit{GPS usage} is estimated as GPS active time per hour and the \textit{Battery Consumption} is estimated in ${mA}$ by the Google Battery Historian tool. 
The \textit{Location Accuracy} was calculated in terms of distance covered in ${meters}$.
Initially, we used the Google Fit application to measure the distance.
We have conducted three trials and considered the averaged value of 3.06 kilometers as the benchmark to compare the location accuracy given by the subject application. 
Of the subject applications, \textit{OSMTracker} cannot calculate the distance covered, hence, we have instrumented a popularly known method to measure the distance covered using the location coordinates given by the subject applications. 
The important validity threat in this category is an error in measuring the GPS usage, battery consumption, and location accuracy.
Therefore, we conducted three trials for each \textit{e}GEN and non-\textit{e}GEN versions of subject applications.
The averaged values of three trials have been considered to avoid the error caused in single trials. 
In addition, the subject applications were executed in the controlled environment by disabling all other user applications. 
Hence, we believe that the values measured are from the subject applications and can answer the research questions. 
\subsection*{Internal Validity} Internal validity is a concern when examining causal relationships. 
In this study, there is a possibility of validity threats in deciding the factors affecting GPS usage, battery consumption, and location accuracy. 
We have conducted preliminary controlled experiments with subject applications developed in our lab with varying location-sensing intervals, battery level and battery charging state to mitigate this threat. 
We have conducted random trials with the sample subject application and analyzed the relationship between variables. 
The first two authors have conducted the experiments, and all the authors were involved while analyzing the cause-effect relationship. 
Finally, we have selected \textit{battery drain} as a primary \textit{independent variable} as it affects the location-sensing interval, GPS usage, Battery consumption and location accuracy. 
Consequently, the affected variables such as \textit{sensing interval, GPS usage, battery consumption and distance covered} are selected as \textit{dependent variable}. 
In addition, we have observed a relationship between the \textit{dependent variables}. 
For instance, the \textit{sensing interval} is set to increase for each battery level drop. 
Consequently, we observed a decrease in GPS usage, battery consumption, and accuracy of distance covered. 
Therefore, we verified this causal relationship also in the trials we conducted for the subject application. 
The results show that the mentioned causal relationship has been maintained between the variables.
We believe that we have considered all aspects of causal relationships in this domain which poses less threat to internal validity. 
The other possibility of internal threat is in the selection of subject applications. 
Initially, we searched the apps on F-Droid with keywords such as \textit{gps logging, distance measurement, path tracker, location service, speed meter, location share, live location tracking, etc}.
We believe that the keywords chosen cover all location-based applications such as map navigation, activity monitoring, etc. In addition, we also searched for open source applications through the Google play store.
After getting 49 candidate applications, we have carefully applied inclusion and exclusion criteria. 
The first two authors have primarily worked on applying inclusion and exclusion criteria. 
The third and fourth authors have been involved in resolving the conflicts in the selection of subject applications. 
To the best of our knowledge, we believe that the selected subject applications are more suitable representative subject applications for evaluating \textit{e}GEN.

\subsection*{External Validity} The external validity refers to the generalizability of the results presented in this study. 
There is a possibility of validity threats in terms of the application domain considered in this case study.
In this study, location-based applications have been selected for evaluating the \textit{e}GEN. 
Hence, the results presented in this study is only can be generalized for location-based applications.
However, the results also can be generalized to the smartphone apps if it uses location as one of the important contexts.
The energy-saving opportunities presented in this study can be applied to such applications if location-sensing drains a significant amount of battery. 
The second possible external validity threat is the ability of DSL and code generator developed as part of \textit{e}GEN.
The domain model of \textit{e}GEN is specifically designed for location-based applications. 
Hence, the capability of \textit{e}GEN can be generalized only to location-based applications. 
The code generator presented in this paper generated only Java code that can be used for Android applications.
Hence, the generated code cannot be used for iOS or Windows smartphone platforms. 
However, the support for iOS and Windows platforms will be added in the future enhancements to \textit{e}GEN.

\subsection*{Reliability} Reliability refers to the extent to which the presented case study is repeatable by other researchers. 
The first repeatability issue would be possible in specifying self-adaptive location-sensing policies and generating code. Therefore, to mitigate this threat, we have uploaded the \textit{e}GEN source code to GitHub\footnote{\url{https://github.com/Kowndinya2000/egen}}, which can be downloaded and used by other researchers.
The second threat in this category is instrumenting the subject applications. 
We have uploaded the instrumented subject applications in the replicable package to mitigate this threat, which is available in the replication package\footnote{\url{https://github.com/Kowndinya2000/egen/tree/master/subject-applications}}. 
The other researchers can build the shared source code and conduct the experiments on their testing device. 
There may be situations where the source code building might result in build errors and failure to compile the .apk files. To address this issue, we have also shared the .apk for both \textit{e}GEN and non-\textit{e}GEN versions of the subject application. 
As we conducted controlled experiments, there is a high possibility of getting deviated results when other researchers repeat the same experiment. 
Therefore, to verify the results, we have shared the bug reports\footnote{\url{https://github.com/Kowndinya2000/egen/tree/master/bug-reports}} that contain the GPS usage and Battery consumption data. 
The bug reports can be uploaded to the Google Battery Historian tool, and presented results can be verified. 
\section{Related Work}\label{S:related}
The related works are classified under two categories: (1) DSL for design-time specification of energy-related properties, (2) Model-driven development of Android apps. 
In this section, the related works found from the literature are summarized and compared with the \textit{e}GEN.
\subsection{DSL for Energy Specification}
We found only one DSL-based approach, namely Energy Estimation Language (EEL) \cite{beziers2020annotating} that considered explicit specification of energy-related properties at design time to the best of our knowledge. 
The authors have used energy estimation formulas to annotate the xDSLs to predict their energy consumption. 
In this approach, during design time itself \textit{energy specialists} can define Energy Estimation Models (EEM) for xDSL and predict how much energy it can consume at run-time for different run-time platforms. 
EEL is a textual domain-specific language written using Xtext language development platform. 
The authors have conducted a case study on AurdinoML models to evaluate this approach. 
The results show that the estimation model produces an estimation error of 4.9\%, between 0.4\% and 17.1\%.
EEL is related to \textit{e}GEN in terms of proposing a textual domain-specific modeling language for explicit specification energy-related properties. 
However, it is different from \textit{e}GEN as it targets cyber-physical systems while \textit{e}GEN target Android apps. 
\textit{e}GEN comes with a DSL and code generator for specifying energy-saving opportunities. 
In contrast, EEL comes with the only DSL for specifying run-time energy consumption, and it doesn't have code generation support.

\subsection{Model-driven development of Android apps}
\subsubsection{Existing Approaches}
MD\textsuperscript{2} \cite{heitkotter2013cross} is an approach for developing mobile apps with model-driven development methods.
It consists of a domain-specific language to specify the data-driven business apps. 
It also contains a code generator for generating native Android and iOS code. 
The language and code generated by MD\textsuperscript{2} follow the Model-View-Controller pattern. 
The \textit{Model} component allows the developers to define the application's data model. 
The \textit{View} component helps in describing the user interface and its elements. 
The \textit{Controller} component aids to describe the user interaction and events associated with the apps. 
The DSL was defined with Xtext\footnote{\url{https://www.eclipse.org/Xtext/}}, and Xtend\footnote{\url{https://www.eclipse.org/xtend/}} defines the code generator. 
The recent version of MD\textsuperscript{2} \cite{heitkotter2015extending} includes the following capabilities: \textit{device-specific layout, extended control structures, and offline computing}.  

Xmob \cite{le2013yet} is a platform-independent DSL for creating mobile applications for multiple platforms. 
It is developed with three sub-languages (Xmob-data, Xmob-ui, Xmob-event) to follow the MVC pattern. 
The \textit{Xmob-data} helps the developers to specify the way retrieving form database, web service, or other data sources. 
The \textit{Xmob-ui} helps the developers to describe the UI elements such as widgets, forms, buttons, etc. 
The \textit{Xmob-event} helps the developers to link the user interfaces and data sources. 
Xmob involves model-to-model transformation and model-to-text transformation to generate the source code of the desired platform. 
The model-to-model transformation converts the platform-independent model to a platform-specific model. 
The model-to-text generates the source code corresponding to the elements in a platform-specific model. 
Xmob uses Xtext for language definition, Kermeta for model-to-model transformation, and Xpand for the code generator.

ADSML \cite{jia2015approach} is an adaptive domain-specific modeling language for native mobile app development. 
It relies on meta-model extraction, meta-model elevation, meta-model alignment, and meta-model unification to create target apps for the Android and iOS platforms. 
The \textit{meta-model extraction} phase extracts the platform-specific meta-models from the targeted platforms native APIs. 
The \textit{meta-model elevation} phase abstracts the platform-specific API models and select the sub-set for further analysis.
The \textit{meta-model alignment} phase find out the similar meta-model elements among different platforms.
Finally, the \textit{meta-model unification} phase creates the platform-independent DSL from the platform-specific models identified in the previous phase. 
The current implementation of ADSML does not have the support for code generation. 

DSL-Comet \cite{vaquero2017active} is the active DSL that targets a smart city or IoT applications.
It primarily runs on mobile devices to tag the location and contextual information on the model elements created by DSL-Comet. 
The DSL-Comet includes Open, Geo, and Contextual DSLs to form an active DSL.
The \textit{Open DSLs} interact with external APIs to retrieve the information related to model elements.
The \textit{Geo DSLs} render the models on the map interface to tag the current location on the models associated with geo-services. 
The \textit{Contextual DSLs} are context-aware and helps to re-organize the model after encountering the contextual changes. 
It has iOS and Eclipse-based editors that permits the users to model either on the mobile or desktop. 
The iOS editor stores the models in JSON format, and the Eclipse-based editor stores the models in XML format. 
The DSL-Comet does not have a code generator to generate source code for the targeted platform. 

Rapid APPlication Tool (RAPPT) \cite{barnett2019supporting} aids the developers in specifying the characteristics of mobile applications using domain-specific visual language and textual language.
Initially, the developers can use visual language to specify the high-level architecture and the number of screens with navigation. 
The developers can then use textual language to add more specific information, such as data schema, authentication, web service, etc., to define the app. 
The model-to-model transformation then takes place to convert the app model to the Android model, where high-level specification will be transferred to Android-specific elements such as classes, activities, fragments, etc. Finally, the RAPPT generates the source code from the Android model that resembles the developer's written code. 
The generated code produces the working prototype, and developers need to add the business logic to deliver the working application. 

MoWebA Mobile \cite{nunez2020model} is a model-driven approach covering the mobile apps' data layer.
This approach mainly defines the data source of application to develop offline access to business applications in case of network connectivity issues. 
This approach consists of three phases: (1) Problem Modeling, (2) Solution Modeling, and (3) Source Code Definition.
The \textit{problem modeling} phase uses the Computational-independent Model (CIM) and Platform-independent Model (PIM).
The \textit{solution modeling} phase uses the Architecture-specific Model (ASM) to specify the architectural requirements.
It uses UML profiles to create Platform-independent models and EMF to convert the PIM to ASM. 
Finally, it uses Acceleo\footnote{\url{https://www.eclipse.org/acceleo/}} to transform models to generate code for developing native applications for Android and Windows platforms.

\begin{table}[]
\caption{Comparison of model-driven development approaches for mobile app development}
\label{tab:mdd-comparison}
\resizebox{\textwidth}{!}{
\begin{tabular}{|c|c|c|c|c|c|c|}
\hline
\textbf{Approach} & \textbf{DSL Type}                                              & \textbf{\begin{tabular}[c]{@{}c@{}}Targeted \\ Platforms\end{tabular}} & \textbf{Domain}                                                      & \textbf{\begin{tabular}[c]{@{}c@{}}Modeling \\ Scope\end{tabular}}              & \textbf{\begin{tabular}[c]{@{}c@{}}Context \\ Awareness\end{tabular}} & \textbf{\begin{tabular}[c]{@{}c@{}}Battery \\ Awareness\end{tabular}} \\ \hline
MD\textsuperscript{2} \cite{heitkotter2013cross}                & Textual                                                        & Android, iOS                                                           & \begin{tabular}[c]{@{}c@{}}Data-driven\\  Business Apps\end{tabular} & \begin{tabular}[c]{@{}c@{}}Data, UI \&\\ User Interaction\end{tabular}          & No                                                                    & No                                                                    \\ \hline
Xmob \cite{le2013yet}             & Textual                                                        & \begin{tabular}[c]{@{}c@{}}Android, iOS\\ \& Windows\end{tabular}      & \begin{tabular}[c]{@{}c@{}}All Mobile\\ Apps\end{tabular}            & \begin{tabular}[c]{@{}c@{}}Data, UI \&\\  Events\end{tabular}                   & No                                                                    & No                                                                    \\ \hline
ADSML \cite{jia2015approach}            & Textual                                                        & Android, iOS                                                           & \begin{tabular}[c]{@{}c@{}}All Mobile\\ Apps\end{tabular}            & All aspects                                                                     & No                                                                    & No                                                                    \\ \hline
DSL-Comet \cite{vaquero2017active}             & Graphical                                                      & N/A                                                                    & \begin{tabular}[c]{@{}c@{}}Smart City\\ Applications\end{tabular}    & \begin{tabular}[c]{@{}c@{}}Business\\ Functions\end{tabular}                    & Yes                                                                   & Partial                                                               \\ \hline
RAPPT \cite{barnett2019supporting}            & \begin{tabular}[c]{@{}c@{}}Graphical\\ \& Textual\end{tabular} & Android                                                                & \begin{tabular}[c]{@{}c@{}}All Mobile\\ Apps\end{tabular}            & Views                                                                           & No                                                                    & No                                                                    \\ \hline
MoWebA Mobile \cite{nunez2020model}             & Graphical                                                      & \begin{tabular}[c]{@{}c@{}}Android,\\ Windows\end{tabular}             & \begin{tabular}[c]{@{}c@{}}Offline \\ Business Apps\end{tabular}     & \begin{tabular}[c]{@{}c@{}}Data Layer \&\\ Network \\ Connectivity\end{tabular} & No                                                                    & No                                                                    \\ \hline
\textit{e}GEN             & Textual                                                        & Android                                                                & \begin{tabular}[c]{@{}c@{}}Location\\ based Apps\end{tabular}        & \begin{tabular}[c]{@{}c@{}}Location\\ Sensing Interval\end{tabular}             & Yes                                                                   & Yes                                                                   \\ \hline
\end{tabular}}
\end{table}

\subsubsection{Comparing approaches}
The comparison of the model-driven development of mobile apps is given in Table \ref{tab:mdd-comparison}. 
The approaches are compared based on the following criteria: \textit{DSL Type, Targeted Platform, Domain, Modeling Scope, Support for Context-awareness, and Support for Energy-awareness}.
As shown in Table \ref{tab:mdd-comparison}, the considered approaches can be broadly classified into two categories, namely, \textit{Textual and Graphical} based on the DSL Type. 
The approaches such as MD\textsuperscript{2} \cite{heitkotter2013cross} , Xmob \cite{le2013yet}, ADSML \cite{jia2015approach} uses the \textit{textual DSL} to specify the app functionalities.
The \textit{graphical DSL} is used in the approaches like DSL-Comet \cite{vaquero2017active}, RAPPT \cite{barnett2019supporting}, and MoWebA Mobile \cite{nunez2020model}.

In this research work, \textit{e}GEN framework adopts the textual DSL for modeling the energy-saving self-adaptive requirements of smartphone applications. 
The \textit{Target Platform} criteria refer to the mobile platform for which the source code generated by the code generator associated with the discussed tools. 
Most of the approaches generate code for multiple platforms such as \textit{Android, iOS, and Windows}. 
The RAPPT \cite{barnett2019supporting} approach considers only the Android platform for code generation.
In this research work, the \textit{e}GEN framework covers only the Android platform, and other platforms will be considered in the future releases of the framework.
The \textit{Domain} criteria refer to the application domain covered by the DSL and code generator. 
As shown in Table \ref{tab:mdd-comparison}, most of the approaches cover all the aspects of mobile apps. 
In contrast approaches such as MD\textsuperscript{2} \cite{heitkotter2013cross},  DSL-Comet \cite{vaquero2017active}, and MoWebA Mobile \cite{nunez2020model} covers the specific application domains. 
Specifically, the MD\textsuperscript{2} \cite{heitkotter2013cross} is for data-driven business apps,  DSL-Comet \cite{vaquero2017active} is for smart city applications, and MoWebA Mobile \cite{nunez2020model} is for business applications with offline access.
As observed from the Table, none of the approaches have considered location-based Android applications. 
Subsequently, in this approach, family of location-based services has been considered as the application domain for DSL and code generator.
The \textit{modeling scope} criteria refer to the elements that can be modeled with the DSL provided in the related approaches.
As shown in Table  \ref{tab:mdd-comparison}, most of the approaches cover the data and UI modeling of mobile apps.
None of the existing approaches have considered modeling the location-sensing of mobile apps. 
On the contrary, this research work's modeling scope covers the location-sensing of mobile apps. 
Finally, the existing approaches have been compared for \textit{self-adaptivity and energy-awareness} support.
As observed from Table \ref{tab:mdd-comparison}, none of the existing approaches has considered the self-adaptivity and energy-awareness of the mobile apps, which is the essential non-functional requirements for the recent generation smartphones.
Therefore, in this research work, the modeling of energy-awareness and self-adaptivity has been considered for location-based Android applications. 

\section{Conclusion and Future Work}
\label{sec:conclusion}
This paper presents the \textit{e}GEN tool for modeling energy-aware self-adaptive behaviors of location-based mobile applications. The domain analyst may use the textual DSML to specify the energy-saving adaptation plans. The developer may use the generated battery-aware code in the existing repositories. The preliminary evaluation presented in this paper shows that the instrumented code shows a considerable reduction in battery consumption for the trials that last for 12-14 minutes. Hence, we believe that the \textit{e}GEN generated code might show significant battery savings when the apps are used for a longer time in real-world scenarios.
Therefore, writing battery-aware code might help developers to reduce unwanted battery consumption in location-based Android applications.

Currently, \textit{e}GEN grammar covers only GPS and does not cover other smartphone sensors for location-sensing. We plan to cover accelerometer and magnetometer in the next version. We also intend to perform controlled experiments at a large scale on greater number of android applications to device a catalog of adaptation policies for different categories of android applications. This also would enable us to strengthen the further versions of \textit{e}GEN.     
\color{black}
The \textit{e}GEN generated code cannot be used for iOS or Windows smartphone platforms. However, the support for iOS and Windows platforms will be added in the future enhancements to \textit{e}GEN. \textit{e}GEN has been evaluated by us through controlled experiments. In future, we have plans to evaluate it with mobile app developers for its usability and completeness.

\begin{acks}
This publication is an outcome of the R\&D work undertaken project under the Visvesvaraya Ph.D. Scheme of Ministry of Electronics \& Information Technology, Government of India, being implemented by Digital India Corporation.
\end{acks}

\bibliographystyle{ACM-Reference-Format}
\bibliography{sample-base}

\end{document}